\documentclass[prl,twocolumn,showpacs]{revtex4}
\usepackage{tabularx}
\usepackage{graphicx}
\usepackage{bm}
\usepackage{amsfonts}
\usepackage{amsmath}
\usepackage{exscale}
\usepackage{amsbsy}

\def\prl{Phys. Rev. Lett.}

\def\be{\begin{equation}}
\def\ee{\end{equation}}
\def\ba{\begin{eqnarray}}
\def\ea{\end{eqnarray}}

\begin{document}

\title{{\it d}-wave to {\it s}-wave to normal metal transitions in disordered superconductors}

\author{B. Spivak}
\affiliation{Department of Physics, University of Washington, Seattle WA 98195, USA}
\author{P. Oreto, S. A. Kivelson}
\affiliation{Department of Physics, Stanford University, Stanford CA  94305, USA}

\begin{abstract}
We study suppression of superconductivity by disorder in  {\it d}-wave
superconductors, and predict the existence of
(at least) two sequential low temperature transitions
 as a function of increasing disorder: a {\it d}-wave to {\it s}-wave, and then an s-wave to metal transition. This is a universal property of the system
 which is independent of the sign of the interaction constant in the {\it s}-channel.
\end{abstract}
\maketitle

Generally the order parameter in superconductors is a function of two coordinates
and two spin indices.
Classification of possible superconducting phases in crystalline materials
was done in  \cite{GorkovVolovik,sigrist}.
A majority of low-$T_{c}$ crystalline superconductors have a singlet order parameter with {\it s}-wave symmetry. It does not change its sign under rotation, and in the isotropic case can be approximated
by a complex number $\Delta^{s}({\bf r})=\Delta({\bf r}, {\bf r})$.
 However, over the last decades a number of
 superconductors have been discovered in which the order parameter  changes
  sign under rotation.
 A notable example is HTC superconductors, where
  in the absence of disorder the order parameter has singlet {\it d}-wave symmetry \cite{harlingen,Kirtley}:
 $\Delta({\bf r}-{\bf r}')$ changes
  sign under rotation by $\pi/2$,  and consequently
  $\Delta({\bf r},{\bf r})=0$. This means that the Fourier transform $\Delta({\bf k})$ changes its sign under a
$\pi/2$ rotation as well, as  is shown schematically by the rosettes in Fig.1.
Since the sign of $\Delta({\bf k})$ in crystalline {\it d}-wave superconductors
depends on the direction of the wave vector ${\bf k}$, they are  much more sensitive to disorder than {\it s}-wave superconductors: at temperature $T=0$, {\it d}-wave superconductivity gets destroyed  when the electron mean free path $l$ is of the order of the zero temperature coherence length in a pure superconductor,
 $ l\sim l_0=1.78 \xi_o \gg 1/k_{F}$.
Here $k_{F}$ is the Fermi wavelength.
This is in contrast with the case of {\it s}-wave superconductors, where according to the Anderson theorem the superconductivity is destroyed  at much higher level of disorder, when $l\sim 1/k_{F}$.
The fate of the {\it d}-wave superconductors at $l<\xi_{0}$ depends on the sign
 of the interaction constant $\lambda_{s}$ in the s-wave channel. If  the interaction $\lambda_{s}$ in the s-wave channel is attractive,
 but weaker than the attraction in the {\it d}-wave channel $|\lambda_{s}|<|\lambda_{D}|$,  then at weak disorder,
 ($l > \xi_{0}$), the superconducting order parameter has d-wave symmetry, while at $l<\xi_{0}$ the disorder destroys the {\it d}-wave
  superconductivity and the system undergoes a phase transition into an s-wave superconducting state. (See, for example, \cite{Vojta}).

In this article we consider a more interesting case, in which the interaction in the {\it s}-channel is repulsive at strong enough disorder
$1/k_{F}\ll l\ll \xi_{0}$  the system is in normal state. We predict  at least two low- temperature phase transitions:  a {\it d}-wave to {\it s}-wave, and then an {\it s}-wave to normal metal transition. Qualitatively the phase diagram of disordered {\it d}-wave superconductors is
shown in Fig.1.
Let us first discuss the definition of
{\it s}- and {\it d}- symmetries in bulk disordered systems.
 Before averaging over random realizations of disorder, the system does not possess any particular spacial symmetry at all.
However in bulk samples, the symmetry is restored upon configuration averaging.
We can think of several different definitions of the global
symmetry of the order parameter:
 a) An operational definition
 is provided by the result of a phase sensitive experiment, such as the corner SQUID experiment, for example, \cite{Kirtley,harlingen}.
 b) The quantity
 $\overline{\Delta({\bf r},{\bf r}'})$ can be characterized as having {\it d}-wave or {\it s}-wave symmetry.  Here the over-line stands for the averaging over the sample volume.
  c) A  globally {\it s}-wave component of the order parameter can be defined in terms of the
  local {\it s}-component of the anomalous Green function
  ${\cal F}({\bf r}={\bf r}')\equiv {\cal F}^{(s)}({\bf r})$.  If we define $P_{\pm}$ to be  the volume fraction
  of a sample
where $F^{(s)}({\bf r})$ has a positive or negative sign, respectively, then
the system has an s-wave component if
$(P^{+}-P^{-})\neq 0$.
These definitions may be not equivalent under all circumstances.  However,
for the most part, we will deal with
  the interval of  parameters
  in which all these definitions are approximately interchangeable.

It is important to realize that it is  inevitable near criticality to have a
situation in which the local pairing in disordered superconductors is ``{\it d}-wavelike'' and yet the global superconductivity has  s-wave symmetry.
The  {\it d}-wave to {\it s}-wave transition can be understood at the mean field level.
The electron mean free path is an average characteristic of disorder.
Let us introduce a "local" value of the mean free path $l({\bf r})$ averaged over
a size of order $\xi_{0}$.
In the region of parameters where {\it d}-wave superconductivity is sufficiently suppressed by disorder,
the spatial dependence of the order parameter can be visualized as a system of superconducting
puddles with anomalously large values of the order parameter, which are connected by
Joshepson links through non-superconducting metal.
The superconductivity inside the puddles may be enhanced because  either the electron interaction constant, or the mean
 free path in the puddles (or both) may be larger than their  average values.

Let us assume that the distance between the puddles is larger than both their size and the mean free path. In this case the system is already in a state with the "global {\it s}-wave" symmetry.
Its origin is illustrated qualitatively in Fig.2,  where a  system of superconduting puddles of arbitrary shape embedded into a metal is shown.
 The order parameter inside the puddles has {\it d}-wave symmetry, and the orientation of the gap nodes is assumed to be pinned by the
crystalline anisotropy.   In a {\it d}-wave superconductor, in addition to an overall phase of the order parameter, there is an arbitrary
 sign associated with the internal structure of the pair wave function.  Specifically, we adopt a uniform phase convention such that
  when the phase of the order parameter $\phi_i= 0$, this implies $\Delta({\bf r}, {\bf r}')$ in puddle $i$ is real and has its
  positive lobes along the $y$ axis and its negative lobes along the $x$ axis.

 The inter-puddle Joshepson coupling originates from the proximity effect in the normal metal. It is characterized by the
 anomalous Green function ${\cal F}({\bf r,r'})\equiv  F({\bf r,r'}, t=t')$,
  which is connected to $\Delta({\bf r},{\bf r}')$ by the interaction constant.
  Due to the lack of symmetry at the boundary of a puddle, an
 {\it s}-wave component ${\cal F}({\bf r}={\bf r}')={\cal F}^{(s)}({\bf r})\neq 0$ of the anomalous Green function is generated in the neighboring metal.
 At a distance from the superconductor-normal metal boundary larger than
 the elastic electron mean free path  the anomalous Green function  becomes isotropic. In other words, only the {\it s}-component ${\cal F}({\bf r}={\bf  r'})={\cal F}^{(s)}({\bf r})$ survives. It is this component that propagates between far separated puddles and determines the Joshepson coupling.

 The sign of ${\cal F}^{(s)}({\bf r})$
 at a normal metal-superconductor boundary,  is determined
by the sign of the {\it d}-wave order parameter in the {\bf k}-direction perpendicular to the boundary. Therefore it changes along the boundary of a puddle.

At a distance from an individual i-th puddle larger than its size and smaller than the distance between the puddles the quantity
  ${\cal F}^{s}({\bf r})$ has a sign $\eta_{i}=\pm 1$, which  depends on the shape of the i-th puddle. This point is illustrated in Fig.2a, where the sign of the anomalous Green function is positive in hatch-marked areas, and negative outside of these areas.

 If the distance between puddles is larger than their size, the sign of the
Joshepson coupling energy $E_{Jos}$ is determined by a product $\eta_{i}\eta_{j}$,
 \begin{equation}
 E_{Jos}=\sum_{i\neq j} \eta_{i}\eta_{j}J^{(s)}_{ij}\cos(\phi_{i}-\phi_{j}).
 \label{MFdsBigR}
 \end{equation}
 Here indexes $i,j$ label puddles,  $J^{(S)}_{ij}>0$.
Eq.~\ref{MFdsBigR} represents the Mattis model, which is well known  in the theory  of spin glasses \cite{mattis}.
   The ground state of this model corresponds to
\begin{equation}
\cos (\phi_{i})= - \eta_{i} .
\end{equation}
 Thus the distribution of $\cos(\phi_{i})$ between puddles looks completely random as it is shown in Fig. 2a.
 However the system is not a glass because it's ground state has a hidden symmetry.
 In other words if the distances  between puddles are bigger than the characteristic size of the puddles, $R$, the  Josephson
  coupling between  puddles inevitably favors globally {\it s}-wave superconductivity, even though the order parameter on each
   puddle looks locally {\it d}-wave -like.  It is obvious that at a high concentration of puddles, the order parameter  in the ground state has global
   {\it d}-wave symmetry (See Fig. 2b.).

At intermediate distances, the situation is more complicated.
Areas with different signs of ${\cal F}^{(s)}({\bf r})$  mix in a random fashion.  We
argue that the most important
aspects of this complex situation can be  modelled by adding to the right hand side of Eq.~\ref{MFdsBigR} a term

\begin{equation}
\sum_{i\neq j} J^{(d)}_{ij}\cos(\phi_{i}-\phi_{j}),
\end{equation}
where $J_{ij}^{(d)}>0$ characterizes the strength of the exchange interaction between the {\it d}-wave components
of the order parameter. Typically, at small $|{\bf r}_{i}-{\bf r}_{j}|$,
$J_{ij}^{(d)}>J_{ij}^{(s)}$, but
at large $|{\bf r}_{i}-{\bf r}_{j}|$ the coupling strength $J_{ij}^{(s)}$ decays more slowly than $J_{ij}^{(d)}$. Here ${\bf r}_{i}$ are coordinates of the puddles.
Thus it is likely that in this intermediate region the system may exhibit spin glass features and/or coexistence of
 {\it d}-wave and {\it s}-wave ordering. In this article, however,  we will not
further explore this fascinating but complex aspect of this problem.

To quantify the picture presented above  one has to compute the Josephson coupling between a pair of far separated puddles. Since the time that it takes for electrons to travel between puddles is shorter than the characteristic time of  fluctuations of the order parameter on individual puddles,  one can calculate  $J^{(s)}_{ij}$
using the mean-field  Usadel equation for the configuration-averaged anomalous Green function
$\langle  F_{\epsilon}^{(s)}({\bf r})\rangle\equiv -i\sin \theta(\epsilon, {\bf r})$
 in the metal,
\begin{equation}
\frac{D_{tr}}{2}\partial^{2}_{{\bf r}}\theta(\epsilon, {\bf r})+i\epsilon \sin \theta(\epsilon, {\bf r}) =0.
\label{UnadelEq}
\end{equation}
Here $D_{tr}$ is the transport diffusion coefficient of electrons in the metal, ${\cal F}_{\epsilon}^{(s)}({\bf r})$ is the Fourier  transform of $ F^{(s)}[{\bf r}, (t-t'))]$, $\Delta^{(s)}=\lambda^{(s)}{\cal F}^{(s)}({\bf r})$, and
the brackets $\langle ...\rangle$ indicate averaging over random scattering potential between the puddles at a given shape of the puddles.
The only, but crucial difference with the conventional case of s-n junctions (See, for example, \cite{Usadel,ramer}), is the boundary
conditions for Eq.~\ref{UnadelEq} at the normal-superconductor surface, which determine the sign of $\eta_{i}$.

For the case when the size of the puddle is larger than the coherence length and the Andreev reflection on the puddles is effective
 the boundary conditions for Eq.\ref{UnadelEq} on the d-n boundary have been derived in Ref. \cite{Nazarov}.
Since the relevant energy for computing the Josephson coupling, $\epsilon\approx D_{tr}/|{\bf r}_{i}- {\bf r}_j|^{2}$, is much smaller than the
value of the order parameter in the puddles,
  the boundary condition for $\theta({\bf r},\epsilon)$ is independent of $\epsilon$ and  depends only on the angle
between the unit vector parallel to the direction of a gap node ${\bf \hat n}_{\Delta}$ and a unit vector, ${\bf \hat n}({\bf r})$, normal
to the boundary at  point ${\bf r}$ at the surface, :
$
\theta_{s}(\epsilon, {\bf r})=f[\alpha({\bf r})], \ \ \  \sin[\alpha({\bf r})]\equiv{\bf \hat n}({\bf r})\cdot{\bf \hat n}_{\Delta}.
\label{boundCond}
$
Here $f(\alpha)$ is a smooth, approximately odd and periodic function, $f(\alpha)\approx -f(-\alpha) $, $f(\alpha) \approx f(\alpha+\pi)$,
  which grows from $f(\alpha)\approx 0$ at $\alpha=0$, to $f(\alpha)\approx \pm \zeta$ for  $\alpha=\pi/4$, where $\zeta\sim 1$.
Solving Eq.~\ref{UnadelEq} with  these boundary conditions,   and using the standard procedure
 of calculation of the Joshepson energy we get
\begin{eqnarray}
J^{s}_{ij}\sim C\frac{V}{|{\bf r}_{i}-{\bf r}_{j}|^{D}}\exp(-\frac{|{\bf r}_{i}-{\bf r}_{j}|}{L_{T}})\nonumber \\
\eta_{i} = {\rm sign}\left\{ \int_{i} d s f(\alpha) \right\}
\label{J_ijpm1}
\end{eqnarray}
 and $C\sim {G}_{eff}\frac{D_{tr}}{\bar{R}^{2}}$, $V$ is the puddle volume,  the integral is taken over the surface of the $i_{th}$ puddle, and $G_{eff}$ is the conductance of a metal of a size of order of the size of the superconducting puddle.
In this case the magnitude of the {\it s}-component of the order parameter generated at the superconductor-normal metal boundary is of order of
the magnitude of the {\it d}-wave component. Thus it is not surprising that  the value of $J_{ij}^{(s)}$
in Eq.~\ref{MFdsBigR} turns out to be of the same order as  in the case of SNS junction.

If the distribution function of the mean-free paths is unbounded,
and with certain probabilities one can find arbitrary large values of $l({\bf r})$, the mean field superconducting solution  always exists.
However, if the puddle concentration is small enough,  the transition from the state with global {\it s}-wave symmetry to the normal metal
is triggered by a competition between the inter-puddle Joshepson coupling energy and the thermal (or quantum) fluctuations.
Thermal fluctuations destroy the coherence between two puddles when $J_{ij}\sim kT$, which gives us an expression for the critical temperature  $T_{c}$ of the {\it s}-wave superconductor-metal transition
 \begin{equation}
 T_{c}\sim \frac{C V}{R^{D}},
 \label{T_{c}}
 \end{equation}
where $R$ is the inter-puddle distance.
 \begin{center}
\begin{figure}
\includegraphics[scale=0.3, bb=43 406 444 581]{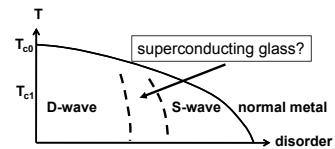}
\caption{Schematic phase diagram for the case when {\it d}-wave superconducting
state  is destroyed as a function of increasing disorder strength.}
\label{fig1}
\end{figure}
\end{center}

 We would like to stress that the existence of the {\it s}-wave superconducting phase is a generic property of the system because the long-range nature of the decay of Eq.~\ref{J_ijpm1} ensures that near the superconductor-normal metal transition and at small enough
temperatures the superconducting puddles are separated by a  distance larger than their size.

In principle, the situation described above can be realized when grains of {\it d}-wave superconductors are embedded into a normal metal artificially.
In random systems the critical point can
be identified by finding
 the set of ``optimal puddles" which lie on the critical links of ``the
percolating cluster".
In this case the properties of the {\it s}-wave phase and the dependence of the critical
temperature $T_{c}$ on the parameters of the system depends on details of the distribution function of the disordered potential.
 To illustrate the situation we consider here a simple model where the mean free part
$l({\bf r})$ is a random function of coordinates with a Gaussian distribution
characterized by an average ${\bar l}$, a variance $\sigma l_{0}$, and a correlation length which is of order $\xi_{0}$. To be concrete, we consider the
2D case.
Then the distance between the puddles becomes of order of their size,
the amplitude of fluctuations of the order parameter becomes of order of the average,
and the system has a transition to the {\it s}-wave state
when $l\sim l_{c1}$ and $T<T_{c1}$
\begin{equation}
\bar{l}_{c1}-l_{0}\sim \sigma^{2} l, \,\,\,\,\,\ T_{c1}\sim \sigma T_{c 0}.
\end{equation}
Here $T_{c0}$ is the critical temperature of a pure {\it d}-wave superconductor.
If $l_{0}-\bar{l}\gg l\sigma^{2}$ the distance between "the optimal puddles" is much bigger than their size.
 We can characterize such puddles by a value of the mean free path $l_{opt}>l_{0}$
averaged over the volume of the puddle. In this case   $\Delta_{opt}\sim \Delta_{0}l_{0}/(l_{opt}-l_{0})^{1/2} \ll \Delta_{0}$, the size of the puddle is of order of the zero temperature coherence length $ \xi_{opt}\sim \xi_{0}l/(l_{opt}-l_{0})^{1/2}\gg \xi_{0}$, and the characteristic distance between the puddles is of order of $ \xi_{opt}\exp[(l_{opt}-\hat{l})^{2}/2\sigma^{2}\bar{l}(l_{opt}-l_{0})]$. Here $\Delta_{0}$ is the magnitude of the order parameter in a pure {\it d}-wave superconductor
at $T=0$.  This expression has a minimum at $(l_{opt}-l_{0})\sim (l_{0}-\bar{l})$, and therefore $R_{opt}\sim  \exp[(l_{0}-\bar{l})/l_{0}\sigma^{2} ]$. Using Eq.~\ref{T_{c}} we get
\begin{equation}
T_{c}\sim T_{c0} \sigma \exp[-\frac{(l_{0}-\bar{l})}{l_{0}\sigma^{2} }].
\label{Tc}
\end{equation}
At very small values of $T_{c}$ the  phase transition between the {\it s}-wave superconducting phase and the normal metal is triggered by
 quantum fluctuations of the order parameter. In this case to determine the point of the
 transition   one has to compare $J_{ij}$ with the "quantum temperature" characterized by the zero temperature superconducting susceptibility $\chi_{i}$ of individual puddles \cite{FeigelmanLarkin,HruskaSpivak,KivelsonSpivakOreto}.
Their finite value  is associated with exponentially rare events of tunneling of the
 superconducting order. Consequently, the value of  $\chi_{i}$ turns out to be exponentially big while the  Joshepson coupling between the puddles decays with the
 inter-puddle distance only as a power law.
   Therefore, a generic feature of the s-n transition is that it takes place  when the distance between optimal puddles is exponentially big \cite{KivelsonSpivakOreto}.
  As a result Eq.~\ref{Tc} holds down to
 exponentially low temperatures.

\begin{figure}
\begin{center}
\includegraphics[scale=0.4, bb=61 279 510 743]{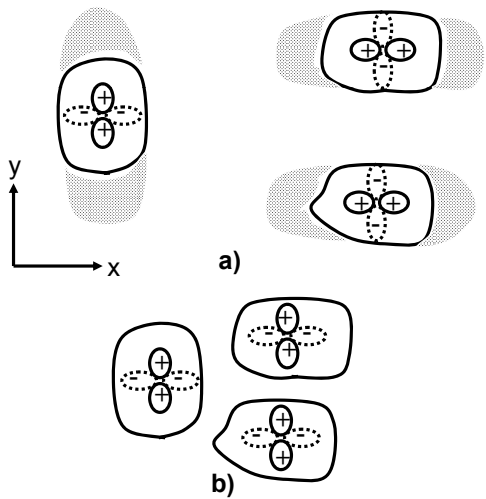}
\caption{A qualitative illustration of the global {\it d}-wave to {\it s}-wave
transition. Solid lines represent boundaries
of {\it d}-wave superconducting puddles embedded into a normal metal. Hatch marked areas
indicate the areas were the {\it s}-wave
component of the anomalous Green function $F^{s}({\bf r}, {\bf r})$ is
positive. Outside these areas $F^{s}({\bf r}, {\bf r})$ is negative.
a) The case of small puddle concentration when the system has {\it s}-wave global symmetry.
b) The case of big puddle concentration when the system has a global {\it d}-wave symmetry.}
\label{fig1}
\end{center}
\end{figure}

The cuprate high-temperature superconductors are the best established example of a {\it d}-wave superconductor.
Here, the critical temperature, $T_c$, is known to vary strongly as a function of the doped hole concentration,
$x$, producing two quantum critical points at which $T_c$ vanishes:  a lower critical doping concentration, $x_{1}$, on the
``underdoped'' side, and an upper critical concentration, $x_2$, on the ``overdoped'' side of the phase diagram.
  On the underdoped side of the superconducting dome,
 with increasing underdoping, these materials frequently appear to undergo a
 superconductor to insulator transition
 \cite{batlogg,ando,taillefer}.  Thus, the present considerations are not applicable.
We assume,  some of the more robust of our findings apply to the cuprates as $T_c \to 0$ with overdoping. There are a number of
 interesting predictions we can make.
1)  There should be a transition from a globally {\it d}-wave to a globally {\it s}-wave superconducting state at a doping concentration $x=x_{2}$.
  (Some evidence of
such a transition may already be present in the experiments of Ref. \onlinecite{swave}.)
2)  In the metallic state with $x > x_2$, the conductivity at low temperature should diverge as $x \to x_2$, the Hall resistance
should vanish, and the Weideman-Franz law should be increasingly strongly violated.

We acknowledge useful discussions with M. Feigelman, D. Fisher, A. Kapitulnik, and  P.
Young.
The research was supported by NSF Grant No. DMR-0704151, and DOE Grant DE-AC02-76SF00515.

\end{document}